\documentclass{article}
\usepackage{graphicx}
\usepackage{amsmath}

\input{tcilatex}

\begin{document}

\title{Three-Point Density Correlation Functions in the Fractional Quantum Hall
Regime}
\author{M. Brownlie and Keith A. Benedict \\
School of Physics and Astronomy, The University of Nottingham, \\
University Park, Nottingham, NG7 2RD, UK.}
\date{27/03/00}
\maketitle

\begin{abstract}
In this paper we consider the three-particle density correlation function
for a fractional quantum Hall liquid. The study of this object is motivated
by recent experimental studies of fractional quantum Hall systems using
inelastic light scattering and phonon absorption techniques. Symmetry
properties of the correlation function are noted. An exact sum-rule is
derived which this quantity must obey. This sum-rule is used to assess the
convolution approximation that has been used to estimate the matrix elements
for such experiments. PACS Numbers: 73.40.Hm, 73.20.Mf, 72.10.Di
\end{abstract}

\section{Introduction}

The fractional quantum Hall effect has been one of the key topics in
condensed matter physics for the last 15 years. While much of the
experimental work on this phenomenon has been focussed on electrical
transport measurements, recently attention has turned to the development of
complementary spectroscopic techniques. In the case of both phonon\cite
{Phonon1, Phonon2} and inelastic light scattering\cite{Pinczuk1, Pinczuk2}
experiments the spectroscopic probe couples to the electronic density of the
system and so naturally interacts with its collective density fluctuations.
The theory of these collective modes was first developed by Girvin,
MacDonald and Platzman (GMP)\cite{GMP} in a manner analogous to that used by
Feynman for the phonon and roton modes of superfluid helium. Unlike the case
of liquid helium, there is no gapless phonon-like behaviour at small wave
vector; the collective modes of the fractional quantum Hall systems are
gapped at all wavelengths: a consequence of the incompressibility of these
states. However, in common with liquid helium, the collective mode
dispersion does have a well defined minimum, hence the quanta of collective
excitations of the fractional quantum Hall states are referred to as
magnetorotons.

One can envision two types of process involved in both the phonon and light
scattering experiments. Firstly a magnetoroton can be created when the
electron liquid absorbs one of the probe quanta (phonon or photon), we refer
to this as type A absorption. Secondly, an existing magnetoroton can be
scattered into a different state by the absorption of a probe quantum, this
we call a type B process. At zero temperature the electron liquid will be in
its ground state (well described by Laughlin's wave function\cite{Laughlin})
and there will be no magnetorotons present, hence only the type A process
can occur. Because the magnetoroton dispersion is gapped at all wave vectors
there is a threshold frequency for this process: no probe quantum can be
absorbed whose energy is less than the minimum gap, $\Delta^{\ast}$. Type B
processes will occur at any non-zero temperature and have no threshold: in
principal, arbitrarily small energies can be transferred to the electron
liquid by these processes. The theory of the Type A processes has been
discussed by He and Platzman \cite{He} for the inelastic light-scattering
case and by our group \cite{Phonon-theory} for the phonon case. These type B
processes will determine the leading finite-temperature corrections to the
zero temperature theory outlined in \cite{Phonon-theory}. The theory of Type
B processes is less well developed but potentially rather important for the
understanding of the phonon experiments because these typically use pulses
containing a black body distribution of phonon energies characterized by a
pulse temperature $T_{\phi}<\Delta^{\ast}$ and use the temperature of the
electron liquid itself to monitor the rate of phonon absorption. At an
arbitrary electron temperature there will be competition between rare type A
processes in which high energy phonons are absorbed and common type B
processes in which low energy phonons are absorbed. In recent time-resolved
experiments\cite{Phonon2} the change in electron temperature is recorded
over the period of time in which the pulse is in contact with the electrons.
At early times the electron liquid is at very low temperatures and so type A
processes will dominate the energy transfer and hence the rate of change of
electron temperature. At later times the electron temperature will be
higher, because of all the earlier type A absorptions and type B processes
will become comparable in determining the energy transfer rate.

The purpose of this paper is to derive a sum rule obeyed by the matrix
elements which control the rates for the type B processes and to use this to
test the best existing approximation scheme for this object: the colvolution
approximation developed for studying the corresponding quantities for
superfluid Helium-4 \cite{Jackson}. As discussed in \cite{MacDonald}, these
matrix elements are also important for the understanding of the effects of
disorder on the form of the magnetoroton dispersion. In all of the
following, natural units in which $l_{c}=\sqrt{\hbar/eB}=1$ and $%
e^{2}/4\pi\epsilon _{0}\kappa l_{c}=1$ (where $\kappa$ is the bulk
dielectric constant of the material in which the 2des is formed) are used.

\section{Magnetorotons}

Girvin, MacDonald and Platzman\cite{GMP} supposed that the low-lying
collective excitations of a fractional quantum Hall fluid could be obtained
in a manner analogous to that used by Feynman in his classic works on
superfluid helium. They proposed that a collective excitation with 2d wave
vector $\mathbf{q}$ would be described by the wave function 
\begin{equation}
\left| \mathbf{q}\right\rangle =\frac{\overline{\rho}_{\mathbf{q}}\left|
\Omega\right\rangle }{\sqrt{\left\langle \Omega\left| \overline{\rho }_{-%
\mathbf{q}}\overline{\rho}_{\mathbf{q}}\right| \Omega\right\rangle }} 
\label{roton}
\end{equation}
where $\left| \Omega\right\rangle $ is the ground state of the system (for
which Laughlin has provided an excellent wave function, at least for the
cases where the Landau level filling $\nu$ is of the form $1/m$ for $m$ odd)
and $\overline{\rho}_{\mathbf{q}}$ is an operator that is derived from the
conventional density operator 
\begin{equation}
\rho_{\mathbf{q}}=\sum_{i=1}^{N}e^{i\mathbf{q\cdot}\widehat{\mathbf{r}}_{i}}
\end{equation}
by projection onto the lowest Landau level. In the notation used by
MacDonald in \cite{LesHouches} 
\begin{align}
\overline{\rho}_{\mathbf{q}} & =\sum_{i=1}^{N}B_{i}\left( \mathbf{q}\right)
\\
B\left( \mathbf{q}\right) & =\mathcal{P}_{0}e^{i\mathbf{q\cdot r}}\mathcal{P}%
_{0}
\end{align}
where $\mathcal{P}_{0}$ acts on the Hilbert space of a single electron to
project out the states within the lowest (spin-polarized) Landau level.
Girvin and Jach\cite{Jach} investigated the properties of these projected
operators and, in particular deduced that 
\begin{equation}
B_{i}\left( \mathbf{k}\right) B_{i}\left( \mathbf{q}\right) =e^{\mathsf{k}%
^{\ast}\mathsf{q}/2}B_{i}\left( \mathbf{k+q}\right)   \label{product}
\end{equation}
where $\mathsf{k}=k_{x}+ik_{y}$. This leads, bearing in mind that the
projection operators for one particle commute with operators associated with
the other, to the commutation relation for the projected density operators 
\begin{align}
\left[ \overline{\rho}_{\mathbf{k}},\overline{\rho}_{\mathbf{q}}\right] &
=\left( e^{\mathsf{k}^{\ast}\mathsf{q}/2}-e^{\mathsf{q}^{\ast}\mathsf{k}%
/2}\right) \overline{\rho}_{\mathbf{k+q}}  \notag \\
& =i\Phi\left( \mathbf{k,q}\right) \overline{\rho}_{\mathbf{k+q}}
\end{align}
where 
\begin{align}
\Phi\left( \mathbf{k,q}\right) & =-i\left( e^{\mathsf{k}^{\ast}\mathsf{q}%
/2}-e^{\mathsf{q}^{\ast}\mathsf{k}/2}\right) \\
& =e^{\mathbf{k\cdot q}/2}2\sin\left( \frac{1}{2}\mathbf{k\wedge q}\right)
\quad.  \notag
\end{align}

The normalization in equation \ref{roton} includes a matrix element which
has the form of a projected static structure factor 
\begin{equation}
\overline{s}\left( q\right) =\frac{1}{N}\left\langle \Omega\left| \overline{%
\rho}_{-\mathbf{q}}\overline{\rho}_{\mathbf{q}}\right| \Omega\right\rangle
\quad.
\end{equation}
GMP showed that this could be related to the true static structure factor 
\begin{equation}
s\left( q\right) =\frac{1}{N}\left\langle \Omega\left| \rho_{-\mathbf{q}%
}\rho_{\mathbf{q}}\right| \Omega\right\rangle
\end{equation}
via the simple relation 
\begin{equation}
\overline{s}\left( q\right) -e^{-q^{2}/2}=s\left( q\right) -1\quad.
\end{equation}
The magnetoroton energy can be written in the form 
\begin{align}
\Delta\left( k\right) & =\left\langle \mathbf{k}\left| \mathcal{H}\right| 
\mathbf{k}\right\rangle -\left\langle \Omega\left| \mathcal{H}\right|
\Omega\right\rangle  \notag \\
& =\frac{\left\langle \Omega\left| \overline{\rho}_{-\mathbf{k}}\left[ 
\mathcal{H},\overline{\rho}_{\mathbf{k}}\right] \right| \Omega\right\rangle 
}{N\overline{s}\left( k\right) }  \notag \\
& =\frac{\overline{f}\left( k\right) }{\overline{s}\left( k\right) }
\end{align}
and GMP showed that the projected oscillator strength, $\overline{f}\left(
k\right) $, can be written explicitly as an integral involving $\overline{s}$
and $\Phi$ only. Hence, obtaining $\overline{s}\left( k\right) $ from the
Laughlin wave function, they found the dispersion relation for these
excitations.

Consider now a bosonic probe quantum (such as a phonon or photon) labelled
by a 3d wave vector $\mathbf{Q}$ which couples to the 2d electrons via a
hamiltonian of the form 
\begin{equation}
H_{int}=\sum_{\mathbf{Q}}M_{\mathbf{Q}}\rho_{\mathbf{q}}\left( a_{\mathbf{Q}%
}+a_{-\mathbf{Q}}^{\dagger}\right)
\end{equation}
where $a_{\mathbf{Q}}$ annihilates a probe quantum with the given wave
vector, $\rho_{\mathbf{q}}$ is the electron density operator described above
(un-projected) and $M_{\mathbf{Q}}$ is some coupling function. For the
detailed forms of $M_{Q}$ relevant to the experimental systems see \cite
{He,Phonon-theory}. The probability per unit time for a magnetoroton
initially in a state $\left| \mathbf{k}\right\rangle $ to absorb a probe
quantum with wave vector $\mathbf{Q}$ and be scattered into the state $%
\left| \mathbf{k+q}\right\rangle $ is given by Fermi's golden rule as 
\begin{equation}
\tau_{\mathbf{k,Q}}^{-1}=\frac{2\pi}{\hbar}\left| M_{\mathbf{Q}}\right|
^{2}\left| \left\langle \mathbf{k+q}\right| \rho_{\mathbf{q}}\left| \mathbf{k%
}\right\rangle \right| ^{2}\delta\left( \Delta\left( \left| \mathbf{k+q}%
\right| \right) -\Delta\left( k\right) -\hbar\omega _{\mathbf{Q}}\right)
\quad
\end{equation}
where $\omega_{\mathbf{Q}}$ is the energy of the probe quantum. This, as
usual with lowest order perturbation theory, factorizes nicely into a part
which depends on the well characterized details of the probe and the
material in which the 2des is formed and a part that only involves the
states of the correlated 2d electrons. The relevant matrix element in the
latter is 
\begin{equation}
\left\langle \mathbf{k+q}\right| \rho_{\mathbf{q}}\left| \mathbf{k}%
\right\rangle =\frac{P\left( \mathbf{k,q}\right) }{\sqrt{\overline{s}\left(
\left| \mathbf{k+q}\right| \right) \overline{s}\left( k\right) }}
\end{equation}
where we have replaced $\rho_{\mathbf{q}}$ with its projected counterpart by
virtue of the idempotence of the projection operators and we define the
three-point correlation function 
\begin{equation}
P\left( \mathbf{k,q}\right) =\frac{1}{N}\left\langle \Omega\left| \overline{%
\rho}_{\mathbf{-k-q}}\overline{\rho}_{\mathbf{q}}\overline{\rho }_{\mathbf{k}%
}\right| \Omega\right\rangle \quad.
\end{equation}
.

\section{Symmetry Properties of the Correlation Function}

We can deduce two symmetry properties of $P\left( \mathbf{k,q}\right) $
straightforwardly. Firstly we can use the fact that $\overline{\rho }_{%
\mathbf{q}}^{\dagger}=\overline{\rho}_{\mathbf{-q}}$ to deduce that 
\begin{align}
\left\{ P\left( \mathbf{k,q}\right) \right\} ^{\ast} & =\left\{ \frac{1}{N}%
\left\langle \Omega\left| \overline{\rho}_{\mathbf{-k-q}}\overline{\rho}_{%
\mathbf{q}}\overline{\rho}_{\mathbf{k}}\right| \Omega\right\rangle \right\}
^{\ast}  \notag \\
& =\frac{1}{N}\left\langle \Omega\left| \overline{\rho}_{-\mathbf{k}}%
\overline{\rho}_{-\mathbf{q}}\overline{\rho}_{\mathbf{k+q}}\right|
\Omega\right\rangle  \notag \\
& =P\left( \mathbf{k+q,-q}\right) \quad.
\end{align}

Secondly we can use the commutation relation for the projected density
operators, as derived by Girvin and Jach\cite{Jach}to simplify 
\begin{align}
P\left( \mathbf{k,q}\right) -P\left( \mathbf{q,k}\right) & =\frac{1}{N}%
\left\langle \Omega\left| \overline{\rho}_{\mathbf{-k-q}}\left[ \overline{%
\rho}_{\mathbf{q}},\overline{\rho}_{\mathbf{k}}\right] \right|
\Omega\right\rangle  \notag \\
& =i\Phi\left( \mathbf{q,k}\right) \frac{1}{N}\left\langle \Omega\left| 
\overline{\rho}_{\mathbf{-k-q}}\overline{\rho}_{\mathbf{k+q}}\right|
\Omega\right\rangle  \notag \\
& =i\Phi\left( \mathbf{q,k}\right) \overline{s}\left( \left| \mathbf{k+q}%
\right| \right) \quad.
\end{align}

\section{Derivation of the Sum Rules}

\subsection{Structure Factor Sum-Rule}

In order to show the basic idea of the method we will begin with a simple
case and prove a sum-rule for the projected structure factor itself. From
the work of GMP \cite{GMP} we know that 
\begin{equation*}
\overline{s}\left( q\right) -e^{-q^{2}/2}=s\left( q\right) -1\qquad. 
\end{equation*}
In first quantized notation we have that, for a system of $N$ particles 
\begin{align*}
s\left( q\right) & =\frac{1}{N}\left\langle \Omega\right| \left(
\sum_{i=1}^{N}e^{-i\mathbf{q\cdot}\widehat{\mathbf{r}}_{i}}\right) \left(
\sum_{j=1}^{N}e^{i\mathbf{q\cdot}\widehat{\mathbf{r}}_{j}}\right) \left|
\Omega\right\rangle \\
& =1+\frac{1}{N}\sum_{i\neq j}\left\langle \Omega\right| e^{-i\mathbf{q\cdot 
}\left( \widehat{\mathbf{r}}_{i}\mathbf{-}\widehat{\mathbf{r}}_{j}\right)
}\left| \Omega\right\rangle
\end{align*}
Hence we have that 
\begin{equation*}
\int\left( s\left( \mathbf{q}\right) -1\right) d^{2}\mathbf{q=}\frac{1}{N}%
\sum_{i\neq j}\left\langle \Omega\right| \int e^{-i\mathbf{q\cdot}\left( 
\widehat{\mathbf{r}}_{i}\mathbf{-}\widehat{\mathbf{r}}_{j}\right) }d^{2}%
\mathbf{q}\left| \Omega\right\rangle 
\end{equation*}
Now 
\begin{equation*}
\int e^{-i\mathbf{q\cdot}\left( \widehat{\mathbf{r}}_{i}\mathbf{-}\widehat{%
\mathbf{r}}_{j}\right) }d^{2}\mathbf{q}=\delta\left( \widehat {\mathbf{r}%
}_{i}\mathbf{-}\widehat{\mathbf{r}}_{j}\right) 
\end{equation*}
so that 
\begin{equation*}
\int\left( s\left( \mathbf{q}\right) -1\right) d^{2}\mathbf{q}=\frac{1}{N}%
\sum_{i\neq j}\left\langle \Omega\right| \delta\left( \widehat {\mathbf{r}%
}_{i}\mathbf{-}\widehat{\mathbf{r}}_{j}\right) \left| \Omega\right\rangle
\qquad. 
\end{equation*}
Inserting a representation of unity in terms of the complete set of $N$%
-particle position eigenstates $\left| \left\{ \mathbf{r}_{i}\right\}
\right\rangle $ gives 
\begin{align*}
\int\left( s\left( \mathbf{q}\right) -1\right) d^{2}\mathbf{q} & =\frac{1}{N}%
\sum_{i\neq j}\int\left( \prod_{i=1}^{N}d^{2}\mathbf{r}_{i}\right)
\left\langle \Omega\right| \delta\left( \widehat{\mathbf{r}}_{i}\mathbf{-}%
\widehat{\mathbf{r}}_{j}\right) \left| \left\{ \mathbf{r}_{i}\right\}
\right\rangle \left\langle \left\{ \mathbf{r}_{i}\right\}
|\Omega\right\rangle \\
& =\frac{1}{N}\sum_{i\neq j}\int\left( \prod_{i=1}^{N}d^{2}\mathbf{r}%
_{i}\right) \delta\left( \mathbf{r}_{i}\mathbf{-r}_{j}\right) \left\langle
\Omega|\left\{ \mathbf{r}_{i}\right\} \right\rangle \left\langle \left\{ 
\mathbf{r}_{i}\right\} |\Omega\right\rangle
\end{align*}
The only configurations that could contribute to this integral are those in
which two particles co-incide. Such configurations, however, have zero
weight because of the Pauli exclusion principle. Hence we conclude that 
\begin{equation*}
\int\left( s\left( \mathbf{q}\right) -1\right) d^{2}\mathbf{q}=0\qquad. 
\end{equation*}
From this we find, straightforwardly that 
\begin{equation*}
\int\overline{s}\left( q\right) d^{2}\mathbf{q}-2\pi\int_{0}^{\infty
}e^{-q^{2}/2}qdq=\int\left( s\left( q\right) -1\right) d^{2}\mathbf{q}=0 
\end{equation*}
or 
\begin{equation*}
\int\overline{s}\left( q\right) d^{2}\mathbf{q}=2\pi\qquad. 
\end{equation*}
Of course, this result can be derived more simply by noting that 
\begin{equation*}
s\left( \mathbf{q}\right) =1+\rho\int e^{-i\mathbf{q\cdot r}}\left( g\left(
r\right) -1\right) d^{2}\mathbf{r}+4\pi^{2}\rho\delta^{2}\left( \mathbf{q}%
\right) 
\end{equation*}
where $g\left( r\right) $ is the radial distribution function. Hence 
\begin{align*}
\int\left( s\left( \mathbf{q}\right) -1\right) d^{2}\mathbf{q} &
=\rho\int\int e^{-i\mathbf{q\cdot r}}d^{2}\mathbf{q}\left( g\left( r\right)
-1\right) d^{2}\mathbf{r}+4\pi^{2}\rho\int\delta^{2}\left( \mathbf{q}\right)
d^{2}\mathbf{q} \\
& =4\pi^{2}\rho\int\delta^{2}\left( \mathbf{r}\right) \left( g\left(
r\right) -1\right) d^{2}\mathbf{r}+4\pi^{2}\rho \\
& =4\pi^{2}\rho g\left( 0\right) \qquad.
\end{align*}
The Pauli principle ensures that $g\left( 0\right) =0$ so that we again
recover our sum-rule.

\subsection{Three Point Correlation Sum-Rule}

Now let us derive our principal result: a sum rule for the three point
correlation function. As shown by MacDonald et al.\cite{MacDonald}, the
three point correlation function can be written in the form 
\begin{equation}
P\left( \mathbf{k,q}\right) =P_{0}\left( \mathbf{k,q}\right) +h^{(3)}\left( 
\mathbf{k,q}\right)
\end{equation}
where 
\begin{align*}
P_{0}\left( \mathbf{k,q}\right) & =e^{-\frac{1}{2}\left| \mathsf{q}\right|
^{2}-\frac{1}{2}\mathsf{k}^{\ast}\mathsf{q}}\left\{ e^{-\frac{1}{2}\left| 
\mathsf{k}\right| ^{2}}+s\left( \left| \mathsf{k}\right| \right) -1\right\}
\\
& +e^{-\frac{1}{2}\left( \mathsf{k+q}\right) ^{\ast}\mathsf{k}}\left(
s\left( \left| \mathsf{q}\right| \right) -1\right) +e^{\frac{1}{2}\mathsf{q}%
^{\ast}\mathsf{k}}\left( s\left( \left| \mathsf{k+q}\right| \right)
-1\right) \qquad.
\end{align*}
Once again, $\mathsf{k=}k_{x}+ik_{y}$ is the complex representation of the
vector $\mathbf{k}$ and 
\begin{equation}
h^{(3)}\left( \mathbf{k,q}\right) =\frac{1}{N}\sum_{l\neq m\neq
n}\left\langle \Omega\left| e^{-i\left( \mathbf{k+q}\right) \cdot \mathbf{r}%
_{l}}e^{i\mathbf{q\cdot r}_{m}}e^{i\mathbf{k\cdot r}_{n}}\right|
\Omega\right\rangle \quad
\end{equation}
is the un-projected three point correlation function of the quantum liquid.

We wish to find a sum rule of the form 
\begin{equation}
\frac{1}{\mathcal{A}}\sum_{\mathbf{q}}P\left( \mathbf{k,q}\right) =F\left( 
\mathbf{k}\right) \quad.
\end{equation}
The term that is likely to cause us difficulty is the un-projected three
point function, however 
\begin{align}
\frac{1}{\mathcal{A}}\sum_{\mathbf{q}}h^{(3)}\left( \mathbf{k,q}\right) & =%
\frac{1}{\mathcal{A}}\sum_{\mathbf{q}}\frac{1}{N}\sum_{l\neq m\neq
n}\left\langle \Omega\left| e^{-i\left( \mathbf{k+q}\right) \cdot \mathbf{r}%
_{l}}e^{i\mathbf{q\cdot r}_{m}}e^{i\mathbf{k\cdot r}_{n}}\right|
\Omega\right\rangle  \notag \\
& =\frac{1}{N}\sum_{l\neq m\neq n}\left\langle \Omega\right| e^{-i\mathbf{k}%
\cdot\mathbf{r}_{l}}\left[ \frac{1}{\mathcal{A}}\sum_{\mathbf{q}}e^{-i%
\mathbf{q}\cdot\mathbf{r}_{l}}e^{i\mathbf{q\cdot r}_{m}}\right] e^{i\mathbf{%
k\cdot r}_{n}}\left| \Omega\right\rangle \quad.
\end{align}
The sum over $q$ is simply a delta function, so that 
\begin{equation}
\frac{1}{\mathcal{A}}\sum_{\mathbf{q}}h^{(3)}\left( \mathbf{k,q}\right) =%
\frac{1}{N}\sum_{l\neq m\neq n}\left\langle \Omega\right| e^{-i\mathbf{k}%
\cdot\mathbf{r}_{l}}\delta^{2}\left( \mathbf{r}_{l}-\mathbf{r}_{m}\right)
e^{i\mathbf{k\cdot r}_{n}}\left| \Omega\right\rangle
\end{equation}
but, as seen above, this must be identically zero, the only configurations
for which the delta-function is non-zero are ones for which the ground state
wave function vanishes by virtue of the Pauli exclusion principle. Hence $%
\frac {1}{\mathcal{A}}\sum_{\mathbf{q}}h^{(3)}\left( \mathbf{k,q}\right) =0$%
\quad$\forall\mathbf{k}$.

The remaining terms are simplified by noting that 
\begin{equation}
e^{-\frac{1}{2}k^{2}}+s\left( k\right) -1=\overline{s}\left( k\right) \quad.
\end{equation}
and by removing possible singularities by writing 
\begin{equation}
s\left( q\right) -1=h\left( q\right) +N\delta_{\mathbf{q,0}}
\end{equation}
where 
\begin{equation}
h\left( q\right) =\rho_{0}\int\left( g\left( r\right) -1\right) e^{i\mathbf{%
q\cdot r}}d^{2}\mathbf{r}
\end{equation}
and $g\left( r\right) $ is the usual radial distribution function of the
electron liquid. Hence we are left with 
\begin{align}
\frac{1}{\mathcal{A}}\sum_{\mathbf{q}}P_{0}\left( \mathbf{k,q}\right) & =%
\frac{1}{\mathcal{A}}\sum_{\mathsf{q}}e^{-\frac{1}{2}\left| \mathsf{q}%
\right| ^{2}-\frac{1}{2}\mathsf{k}^{\ast}\mathsf{q}}\overline{s}\left(
\left| \mathsf{k}\right| \right) +2\rho_{0}e^{-\frac{1}{2}\left| \mathsf{k}%
\right| ^{2}}  \notag \\
& +\frac{1}{\mathcal{A}}\sum_{\mathsf{q}}e^{-\frac{1}{2}\left( \mathsf{k+q}%
\right) ^{\ast}\mathsf{k}}h\left( \left| \mathsf{q}\right| \right) +\frac{1}{%
\mathcal{A}}\sum_{\mathsf{q}}e^{\frac{1}{2}\mathsf{q}^{\ast}\mathsf{k}%
}h\left( \left| \mathsf{k+q}\right| \right)  \notag \\
& =\frac{1}{\mathcal{A}}\sum_{\mathsf{q}}e^{-\frac{1}{2}\left| \mathsf{q}%
\right| ^{2}-\frac{1}{2}\mathsf{k}^{\ast}\mathsf{q}}\overline {s}\left(
\left| \mathsf{k}\right| \right) +2\rho_{0}e^{-\frac{1}{2}\left| \mathsf{k}%
\right| ^{2}} \\
& +2e^{-\frac{1}{2}\left| \mathsf{k}\right| ^{2}}\frac{1}{\mathcal{A}}\sum_{%
\mathsf{q}}e^{-\frac{1}{2}\mathsf{q}^{\ast}\mathsf{k}}h\left( \left| \mathsf{%
q}\right| \right) \quad.
\end{align}
As shown below in the appendix the summations can be carried out in the
limit $N,\mathcal{A}\rightarrow\infty$, $N/\mathcal{A}=\rho_{0}$ \ to give 
\begin{align}
\frac{1}{\mathcal{A}}\sum_{\mathbf{q}}P\left( \mathbf{k,q}\right) & =\frac{%
\overline{s}\left( \left| \mathsf{k}\right| \right) }{2\pi}+2\rho_{0}e^{-%
\frac{1}{2}\left| \mathsf{k}\right| ^{2}}+2e^{-\frac{1}{2}\left| \mathsf{k}%
\right| ^{2}}\left( -\rho_{0}\right)  \label{sum rule} \\
& =\frac{\overline{s}\left( k\right) }{2\pi}\quad.
\end{align}
which is our desired sum-rule.

We can use the symmetry properties derived above to derive a second version
of this. Consider 
\begin{align}
\frac{1}{\mathcal{A}}\sum_{\mathbf{q}}\left( P\left( \mathbf{k,q}\right)
-P\left( \mathbf{q,k}\right) \right) & =i\frac{1}{\mathcal{A}}\sum_{\mathbf{q%
}}\Phi\left( \mathbf{q,k}\right) \overline{s}\left( \left| \mathbf{k+q}%
\right| \right)  \notag \\
& =i\frac{1}{\mathcal{A}}\sum_{\mathbf{q}}\Phi\left( \mathbf{q-k,k}\right) 
\overline{s}\left( q\right) \quad.
\end{align}
It is shown in the appendix that this integral vanishes so that 
\begin{equation}
\frac{1}{\mathcal{A}}\sum_{\mathbf{k}}P\left( \mathbf{k,q}\right) =\frac{%
\overline{s}\left( q\right) }{2\pi}\quad.
\end{equation}
Finally the reality of the right-hand side of \ref{sum rule} allows a
further form of the sum-rule to be written as 
\begin{equation}
\frac{1}{\mathcal{A}}\sum_{\mathbf{q}}P\left( \mathbf{k+q,-q}\right) =\frac{%
\overline{s}\left( k\right) }{2\pi}\quad.
\end{equation}

The basic idea used here could be extended to consider sum-rules for higher
order correlation functions were they to be of interest.

\section{Assessment of the Convolution Approximation}

In their work MacDonald et al. \cite{MacDonald} estimated the small $q$
behaviour of $P\left( \mathbf{k,q}\right) $ by using the convolution
approximation\cite{Jackson} for the 3-particle distribution function 
\begin{equation*}
n^{\left( 3\right) }\left( \mathbf{r,r}^{\prime}\mathbf{,r}^{\prime\prime
}\right) =\sum_{l\neq m\neq n}\left\langle \Omega\right| \delta\left( 
\mathbf{r-r}_{l}\right) \delta\left( \mathbf{r}^{\prime}-\mathbf{r}%
_{m}\right) \delta\left( \mathbf{r}^{\prime\prime}\mathbf{-r}_{n}\right)
\left| \Omega\right\rangle 
\end{equation*}
which can be written as 
\begin{align*}
n_{c}^{\left( 3\right) }\left( \mathbf{r,r}^{\prime}\mathbf{,r}%
^{\prime\prime}\right) & =\rho^{3}\left\{ 1+h\left( r-r^{\prime}\right)
+h\left( r^{\prime}-r^{\prime\prime}\right) +h\left( r^{\prime\prime
}-r\right) \right\} \\
& +\rho^{3}\left\{ h\left( r-r^{\prime}\right) h\left( r^{\prime
}-r^{\prime\prime}\right) +h\left( r^{\prime}-r^{\prime\prime}\right)
h\left( r^{\prime\prime}-r\right) +h\left( r^{\prime\prime}-r\right) h\left(
r-r^{\prime}\right) \right\} \\
& +\rho^{4}\int h\left( r-R\right) h\left( r^{\prime}-R\right) h\left(
r^{\prime\prime}-R\right) d^{2}\mathbf{R}\qquad.
\end{align*}
The three-point function, $h^{\left( 3\right) }\left( \mathbf{k,q}\right) ,$
can be written as 
\begin{equation*}
h^{\left( 3\right) }\left( \mathbf{k,q}\right) =\frac{1}{N}\int d^{2}\mathbf{%
r}d^{2}\mathbf{r}^{\prime}d^{2}\mathbf{r}^{\prime\prime }e^{-i\left( \mathbf{%
k+q}\right) \cdot\mathbf{r}}e^{i\mathbf{q\cdot r}^{\prime}}e^{i\mathbf{%
k\cdot r}^{\prime\prime}}n_{c}^{\left( 3\right) }\left( \mathbf{r,r}^{\prime}%
\mathbf{,r}^{\prime\prime}\right) 
\end{equation*}
which gives, using the convolution approximation 
\begin{align*}
h_{c}^{\left( 3\right) }\left( \mathbf{k,q}\right) & =N^{2}\delta_{\mathbf{%
k,0}}\delta_{\mathbf{q,0}}+N\delta_{\mathbf{k,0}}h\left( q\right) +N\delta_{%
\mathbf{q,0}}h\left( k\right) +N\delta_{\mathbf{k+q,0}}h\left( k\right) \\
& +h\left( k\right) h\left( q\right) +\left[ h\left( k\right) +h\left(
q\right) +h\left( k\right) h\left( q\right) \right] h\left( \left| \mathbf{%
k+q}\right| \right) \qquad.
\end{align*}
This approximation improves on the standard Kirkwood decoupling in that it
correctly gives the $q\rightarrow0$ limit as 
\begin{equation*}
h_{c}^{\left( 3\right) }\left( \mathbf{k,0}\right) =\left( N-2\right) \left(
s\left( k\right) -1\right) . 
\end{equation*}
Since $P_{0}\left( \mathbf{k,q}\right) $ saturates our sum-rule we expect
that 
\begin{equation*}
\frac{1}{A}\sum_{\mathbf{q}}h^{\left( 3\right) }\left( \mathbf{k,q}\right)
=0\qquad. 
\end{equation*}
Using the convolution form gives 
\begin{align*}
\frac{1}{A}\sum_{\mathbf{q}}h_{c}^{\left( 3\right) }\left( \mathbf{k,q}%
\right) & =\rho N\delta_{\mathbf{k,0}}+N\delta_{\mathbf{k,0}}\frac{1}{A}%
\sum_{\mathbf{q}}h\left( q\right) +2\rho h\left( k\right) \\
& +\frac{1}{A}\sum_{\mathbf{q}}\left\{ h\left( k\right) h\left( q\right)
+\left( h\left( k\right) +h\left( q\right) +h\left( k\right) h\left(
q\right) \right) h\left( \left| \mathbf{k+q}\right| \right) \right\}
\end{align*}
Now 
\begin{align*}
\frac{1}{A}\sum_{\mathbf{q}}h\left( q\right) & \rightarrow\rho\int \frac{%
d^{2}\mathbf{q}}{\left( 2\pi\right) ^{2}}\int d^{2}\mathbf{r}e^{-\mathbf{%
q\cdot r}}\left( g\left( r\right) -1\right) \\
& =-\rho
\end{align*}
so that 
\begin{align*}
F\left( k\right) & =\frac{1}{A}\sum_{\mathbf{q}}h_{c}^{\left( 3\right)
}\left( \mathbf{k,q}\right) =\left( 1+h\left( k\right) \right) \frac
{1}{A}\sum_{\mathbf{q}}h\left( q\right) h\left( \left| \mathbf{k+q}\right|
\right) \\
& \rightarrow\left( 1+h\left( k\right) \right) \int\frac{d^{2}\mathbf{q}}{%
\left( 2\pi\right) ^{2}}h\left( q\right) h\left( \left| \mathbf{k+q}\right|
\right) \qquad.
\end{align*}
This final integral requires a specific form for the pair correlation
function, $h\left( q\right) $. For the primary fractional quantum Hall
states ($\nu=1/m$, for odd $m$) this is known from Monte-Carlo studies of
the Laughlin wave function \cite{GMP} which lead to the form 
\begin{equation*}
h\left( q\right) =-\nu e^{-q^{2}/2}+4\nu e^{-q^{2}}\sum_{m}c_{m}L_{m}\left(
q^{2}\right) 
\end{equation*}
where the $c_{m}$ co-efficients are tabulated in \cite{GMP}. We have
estimated $F\left( k\right) $ numerically for the case $\nu=1/3$ and it is
plotted in figure 1, along with a plot of $\overline{s}\left( k\right) /2\pi$
for comparison. Clearly $F\left( k\right) $ is not identically zero and is
not even negligible in comparison to $\overline{s}\left( k\right) /2\pi$.
Hence we deduce that this approximation is deficient. MacDonald et al.\cite
{MacDonald} themselves pointed out that this approximation could well be
unreliable as it does not correctly reflect the particle-hole symmetry $%
\left( \nu\rightarrow1-\nu\right) $ of the system.

\section{Summary and Discussion}

In this short note we have derived a sum rule (with symmetry related
variants) for the static structure factor and the three-point correlation
function that will determine the leading finite-temperature corrections to
the absorption rates of phonons and photons in spectroscopic studies of the
fractional quantum Hall effect. In principle the approach followed would
allow the formulation of similar sum-rules for higher order correlation
functions should such objects ever become relevant to experimental work. The
sum-rule for the three-point function has been used to assess the validity
of the convolution approximation \cite{Jackson} for fractional quantum Hall
systems and has shown that it does not capture all of the physics.

\emph{This work was supported by the EPSRC (UK).}

\appendix

\section{Evaluation of Integrals}

\label{integrals}First of all we need to evaluate the following integral for
later use: 
\begin{equation}
\int_{0}^{2\pi}e^{-ze^{i\phi}}d\phi=\int_{0}^{2\pi}e^{-z\cos\phi}\left\{
\cos\left( z\sin\phi\right) -i\sin\left( z\sin\phi\right) \right\} d\phi
\end{equation}
now \cite{Gradsteyn} 
\begin{align*}
\int_{0}^{2\pi}e^{-z\cos\phi}\cos\left( z\sin\phi\right) d\phi & =2\pi \\
\int_{0}^{2\pi}e^{-z\cos\phi}\sin\left( z\sin\phi\right) d\phi & =0
\end{align*}
so that 
\begin{equation}
\int_{0}^{2\pi}e^{-ze^{i\phi}}d\phi=2\pi\quad\forall z\quad.
\end{equation}
Now we need to evaluate 
\begin{equation}
f_{1}\left( k\right) =\frac{1}{\mathcal{A}}\sum_{q}e^{-\frac{1}{2}\left|
q\right| ^{2}-\frac{1}{2}k^{\ast}q}
\end{equation}
in the limit $\mathcal{A}\rightarrow\infty$, $N\rightarrow\infty$, $N/%
\mathcal{A}=\rho_{0}$ this becomes 
\begin{equation}
\frac{1}{4\pi^{2}}\int_{0}^{\infty}dqqe^{-q^{2}/2}\int_{0}^{2\pi}d\phi e^{-%
\frac{1}{2}kqe^{i\phi}}=\frac{1}{2\pi}\quad.
\end{equation}
Similarly 
\begin{align}
f_{2}\left( k\right) & =\frac{1}{\mathcal{A}}\sum_{q}e^{-q^{\ast}k/2}h\left(
\left| q\right| \right)  \notag \\
& \rightarrow\frac{1}{4\pi^{2}}\int_{0}^{\infty}dqqh\left( q\right)
\int_{0}^{2\pi}d\phi e^{-\frac{1}{2}kqe^{i\phi}}  \notag \\
& =\frac{1}{2\pi}\int_{0}^{\infty}dqqh\left( q\right) \quad.
\end{align}
The final integral can be evaluated quite generally. 
\begin{align}
\int_{0}^{\infty}dqqh\left( q\right) & =\frac{1}{2\pi}\int d^{2}\mathbf{q}%
h\left( q\right)  \notag \\
& =2\pi\rho_{0}h\left( 0\right) \quad.
\end{align}
Now for any spin-polarized fermi system $g\left( r\right) =1+h\left(
r\right) \rightarrow0$ as $r\rightarrow0$ as a consequence of the Pauli
principle, hence $h\left( 0\right) =-1$ and 
\begin{equation}
\frac{1}{2\pi}\int_{0}^{\infty}dqqh\left( q\right) =-\rho_{0}\quad.
\end{equation}

Finally we need to evaluate 
\begin{equation}
\frac{1}{\mathcal{A}}\sum_{\mathbf{q}}\Phi\left( \mathbf{q-k,k}\right) 
\overline{s}\left( q\right) \rightarrow\frac{1}{4\pi^{2}}e^{-k^{2}/2}%
\int_{0}^{\infty}dqq\overline{s}\left( q\right) \int_{0}^{2\pi}d\phi2e^{%
\frac{1}{2}kq\cos\phi}\sin\left( \frac{1}{2}kq\sin\phi\right) \quad.
\end{equation}
As we have already seen the $\phi$ integral vanishes and so, therefore does
the whole expression.

\section{\protect\bigskip Figure Caption}

Figure 1: A plot of the function $F\left( k\right) $ estimated numerically
(full line) with a plot of $\overline{s}\left( k\right) /2\pi$ plotted
(dashed line) for comparison.


\begin{thebibliography}{99}
\bibitem{Phonon1}  C.J. Mellor, R.H. Eyles, J.E. Digby, A.J. Kent, K.A.
Benedict, L.J. Challis, M. Henini, C.T. Foxon, J.J. Harris, Phys. Rev.
Lett., \textbf{74}, 2339 (1995).

\bibitem{Phonon2}  U. Zeitler, A.M. Devitt, J.E. Digby, C.J. Mellor, A.J.
Kent, K.A. Benedict, T. Cheng, Phys. Rev. Lett., \textbf{82}, 5333 (1999).

\bibitem{Pinczuk1}  A. Pinczuk, B. Dennis, L.N. Pfeiffer, K. West, Phys.
Rev. Lett., \textbf{70}, 3983 (1993).

\bibitem{Pinczuk2}  A. Pinczuk, B. Dennis, A.S. Plaut, L.N. Pfeiffer, K.
West, Proceedings of the 12th International Conference on High Magnetic
Fields in Semiconductors, p83 (Singapore: World Scientific) (1997).

\bibitem{GMP}  S.M. Girvin, A.H. MacDonald, P.M. Platzman, Phys. Rev. 
\textbf{B33}, 2481 (1986).

\bibitem{Laughlin}  R.B. Laughlin, Phys. Rev. Lett., \textbf{50}, 1395
(1983).

\bibitem{He}  P.M. Platzman, S. He, Phys. Rev. \textbf{B49}, 13674 (1994).

\bibitem{Phonon-theory}  K.A. Benedict, R.K. Hills, C.J. Mellor, Phys. Rev.
B \textbf{60}, 10984 (1999).

\bibitem{LesHouches}  A.H. MacDonald, in Course 12 of \emph{Mesoscopic
Quantum Physics: Proceedings of the 61st Les Houches Summer School,} ed E.
Akkermans, G. Montambaux, JL Pichard and J. Zinn-Justin, (Amsterdam:
Elsevier) (1994).

\bibitem{Jach}  S.M. Girvin and T. Jach, Phys. Rev. \textbf{B29}, 5617
(1984).

\bibitem{MacDonald}  A.H. MacDonald, K.L. Liu, S.M. Girvin, P.M. Platzman,
Phys. Rev. \textbf{B33}, 4014 (1986).

\bibitem{Jackson}  H.W.\ Jackson and E. Feenberg, Rev. Mod. Phys. \textbf{34}%
, 686 (1962).

\bibitem{Gradsteyn}  I.S.\ Gradsteyn and I.M.\ Ryzhik, \emph{Tables of
Integrals, Series and Products}, formulae 3.931.4 and 3.931.2, 4th edition
ed. and trans. by A. Jeffreys (New York: Academic Press) (1980).
\end{thebibliography}
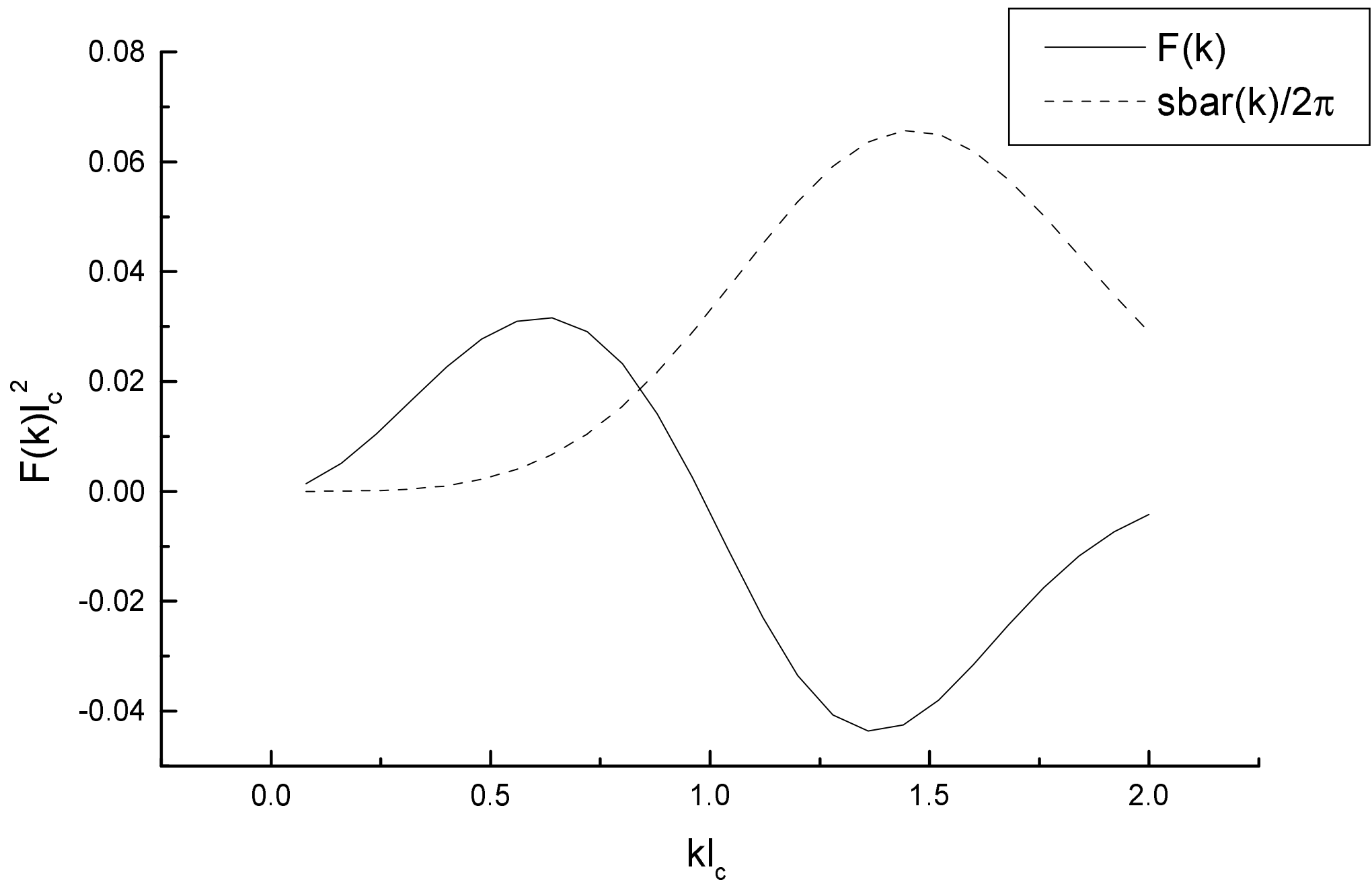
\end{document}